\documentclass[%
reprint, 
groupedaddress,
 amsmath,amssymb,
 amsfonts,
 aps,
 prd,
floatfix
]{revtex4-2}

\usepackage{graphicx}
\usepackage{dcolumn}
\usepackage{bm}
\usepackage{hyperref} 
\usepackage{xcolor}
\usepackage[section]{placeins}
\usepackage{multirow}

\hypersetup{linktocpage=true } 

\begin{document}


\title{No $H_0$ assistance from assisted quintessence}

\author{Vivian I. Sabla}
\email{vivian.i.sabla.gr@dartmouth.edu}
\author{Robert R. Caldwell}%
\affiliation{%
 Department of Physics and Astronomy, Dartmouth College, \\ HB 6127 Wilder Laboratory, Hanover, NH 03755 USA
}%

\date{\today}

\begin{abstract}
Early dark energy, as a proposed solution to the Hubble tension, faces an additional ``why now" problem. Why should dark energy emerge just prior to recombination, billions of years before the onset of cosmic acceleration? Assisted quintessence explains this connection by positing that multiple scaling fields build up over time to drive the present-day cosmic acceleration. In this framework, early dark energy is inevitable. Yet, we show that scaling also leads to the demise of the scenario: the same feature that solves the coincidence problem then spoils a concordance of the Hubble constant inferred from the cosmic microwave background with that from the local distance ladder. The failure of the model offers a novel lesson on the ability of new physics to resolve the Hubble tension.
\end{abstract}

\maketitle

\section{Introduction}

The standard cosmological model, $\Lambda$CDM, provides an extraordinarily good description of a wealth of cosmological observations. Yet, the successes of the model only distract from its inherent flaws and gaps, such as the unknown nature of the dominant dark sector. Further stress on the model has come in recent years, as increased precision in measurements of the cosmic microwave background (CMB), baryon acoustic oscillations (BAO), and Type Ia supernovae have led to a number of tensions between local (or late-time) measurements and inferences based on observations of early-time (pre-recombination) phenomena. The most notable is the discrepancy between values of the Hubble constant, $H_0$, as inferred from the CMB \cite{Aghanim2018} or other independent early-time cosmological probes \cite{Addison2017,Schoneberg2019,Philcox2020}, and as directly measured in the local universe \cite{Riess2018,Riess2019,Wong2019,Huang2019,Pesce2020}. This ``Hubble tension'' can be seen most clearly between the SH0ES measurement of $H_0=74.03\pm1.42$ km/s/Mpc \cite{Riess2019} and the early-universe \textit{Planck} measurement of $H_0=67.4\pm0.5$ km/s/Mpc \cite{Aghanim2018} which differ by $\sim 5\sigma$. Note that some late-universe measurements of $H_0$ \cite{Pesce2020,Huang2019,Freedman2019} are in agreement, within uncertainties, with both CMB inferred values and other late-universe measurements of the Hubble constant. Yet, the $H_0$ tension does not seem to be able to be explained by systematics in either measurement \cite{Efstathiou2013,Addison2015,Aghanim2016,Aylor2018}, suggesting there may be new physics beyond the standard $\Lambda$CDM cosmological model \cite{Hojjati2013,Freedman2017,Feeney2017,Evslin2017}, particularly in the era just prior to recombination \cite{Verde2019,Knox2019}.   

There have been numerous attempts to solve the Hubble tension, focusing on both expansion epochs in question \cite{DiValentino:2021izs}. Possible late-time resolutions include a vacuum phase transition \cite{DiValentino2017,Khosravi2017,Banihashemi2018, Banihashemi2018b,Benevento2020}, modified gravity
\cite{Barreira2014,Umilta2015,Ballardini2016,Renk2017,Belgacem2017,Nunes2018,Desmond2019,Desmond2020}, phantom dark energy \cite{DiValentino2016,DiValentino2017b,Ye:2020btb}, or interacting dark energy \cite{Kumar2016,DiValentino2017c}. Model independent parametrizations of the late-time expansion history also have some success at relieving the tension \cite{Bernal2016,Zhao2017,Vonlanthen2010,Verde2016,Evslin2017}. However, all these late-time resolutions are challenged by tight constraints from late-time observables \cite{Riess2019,DiValentino2017b,DiValentino2017c,Addison2017}, particularly BAO \cite{Beutler2011,Ross2014,Alam2016}. Early time resolutions which modify pre-recombination physics are suggested to be the most likely solutions to the tension \cite{Knox2019}. Many such resolutions have been proposed, including interacting or decaying dark matter \cite{Lesgourgues2015,DiValentino2017d,Bringmann2018,Pandey2019,Audren2014,Blinov2020,Alcaniz2019,Hooper2019}, early modified gravity \cite{Lin2018,Abadi:2020hbr,Zumalacarregui:2020cjh,Braglia:2020iik,Ballesteros:2020sik}, modified neutrino physics \cite{Weinberg2013,Shakya2016,Mortsell2018,Poulin2018,Escudero2019nu}, and early dark energy (EDE) \cite{Karwal2016,Poulin2018ULA,Poulin2018ede,DEramo2018,Smith2019,Lin2019,Agrawal2019,Ivanov:2020ril,Hill:2020osr,Niedermann:2020dwg,DAmico:2020ods,Smith:2020rxx,Sakstein:2019fmf,CarrilloGonzalez:2020oac,Braglia:2020bym}.

EDE, while one of the most promising scenarios, appears fine tuned. Why should dark energy, or a related dark-sector field, emerge near matter-radiation equality at a trace amplitude -- just enough to shift the length scales imprinted into the CMB -- before falling dormant? Not only that, dark energy itself appears fine tuned -- why should it come to dominate so late in the history of the Universe? Surprisingly, both of these issues are addressed in an assisted quintessence scenario \cite{Kim2005}.

In assisted quintessence (AQ), multiple scaling fields are present. None of the fields alone is sufficient to drive cosmic acceleration. But as time progresses, more and more such fields thaw from the Hubble friction and activate, becoming dynamical. Due to the scaling behavior, the fields evolve as a tiny but constant fraction of the background energy density. Eventually, the cumulative effect of the scaling fields is enough to catalyze cosmic acceleration. In this context, given a spectrum of scaling fields, early dark energy and dark energy are inevitable: EDE is just the thaw and activation of a scaling field; dark energy is the cumulative effect of a series of EDE fields.

In this work we assume the true value of $H_0$ is the one implied by the SH0ES measurement of $H_0=74.03\pm1.42$ km/s/Mpc \cite{Riess2019}, and we explore the ability of an individual AQ field to bring the CMB-derived value into concordance. We show that, at the background level, this new component of the energy density seems to be the balm needed to relieve the tension. However, the scaling behavior of the AQ field leaves a significant imprint on the inhomogeneities, ultimately spoiling the concordance and, in fact, exacerbating the tension.

This paper is organized as follows. In Sec.~\ref{sec: model} we introduce the AQ-EDE model as a potential solution to the Hubble tension. We describe the background solution as well as the behavior of linear perturbations. We present the cosmological data used in the MCMC analysis of the model in Sec.~\ref{sec: data and methods}. Results of our parameter estimation, in particular $H_0$, are given in Sec.~\ref{sec: observational constraints}. We conclude our discussion in Sec.~\ref{sec: conclusion} with our main findings. Appendices provide details of the numerical implementation of the model, and extended data analysis results. 

\section{Cosmological Model}
\label{sec: model}

The proposed scenario consists of the standard cosmological model, with dark energy in the form of assisted quintessence. The action is
\begin{equation}
    S = \int d^4x \sqrt{g} \left( \tfrac{1}{2} M_P^2 R + {\cal L}_{M} -\sum_i [\tfrac{1}{2}(\partial\phi_i)^2+V_i(\phi_i) ] \right),
\end{equation}
where ${\cal L}_M$ represents the Standard Model plus cold dark matter, and the index $i$ sums over the contributions of the AQ scaling fields. In the following, we describe the background dynamics, the proposed solution to the Hubble tension, and the behavior of linear perturbations used to evaluate the imprint of the AQ-EDE model on the CMB and the inferred Hubble constant.

\subsection{Background Dynamics}

Tracking fields, proposed as a way to circumvent the cosmic coincidence problem \cite{Steinhardt1999}, have an attractor-like solution leading to a common evolutionary track. For dark-energy tracking solutions, the equation of state $w_\phi$ is a constant, less than or equal to the equation of state of the background fluid $w_B$. Scaling is a special case of tracking where the scaling fields have the same equation of state as the background, $w_\phi = w_B$. For an individual field with potential $V(\phi)$, the capacity for scaling or tracking behavior depends on the quantity \cite{Steinhardt1999}
\begin{equation}
    \Gamma = \frac{ V_{,\phi\phi}V}{(V_{,\phi})^2}
\end{equation}
where $V_{,\phi}=\partial V/\partial\phi$. For convergence to a tracking solution, $\Gamma$ must be nearly constant, in which case the equation of state is
\begin{equation}
    w_\phi \approx \frac{w_B - 2(\Gamma - 1)}{1 + 2(\Gamma-1)}.
\end{equation}
Scaling requires $\Gamma \approx 1$, which implicates an exponential potential.

Exponential potentials arise naturally in higher-dimensional particle physics theories including Kaluza-Klein and string theories, and a variety of super-gravity models. In cosmology, they have mainly been studied within the context of inflation and for a possible role in late-time cosmology. See Refs.~\cite{Ferreira1997,Lucchin1984,Halliwell1986,Shafi1986,Ratra1987,Burd1988,Wetterich1994,Copeland1997,Chang2016} and references therein.

We consider a sequence of exponential potentials of the form 
\begin{equation}
    V_i(\phi_i) = \mu_i^4 e^{-\beta_i \phi_i}.  
\end{equation}
For a single AQ field with this potential, the resulting field evolution,
\begin{equation}
\label{eq: homogeneous KG}
    \ddot\phi + 3 H \dot\phi + V_{,\phi}=0,
\end{equation}
has a well-known exact solution in a background with equation of state $w_B$ \cite{Ferreira1997},
\begin{equation}
   \beta \phi(t) =  \ln\left( \frac{1}{2}\frac{1+w_B}{1-w_B}\beta^2 \mu^4  t^2\right).
   \label{eqn:phi}
\end{equation}
This scaling solution yields an energy density that is a constant fraction of the dominant background
\begin{equation}
    \Omega_\phi(t) = 3(1+w_B)/\beta^2.
\end{equation}
The parameter $\beta$ controls the energy density and is analogous to $f_\text{EDE}$ of Ref.~\cite{Poulin2018ede}. For self-consistency of solution, $\beta^2 > 1/3(1+w_B)$ is required. When $\beta$ is too low, the potential is sufficiently flat that the scalar field will inflate rather than scale. 

\begin{figure}[t]
\includegraphics[width=0.47\textwidth]{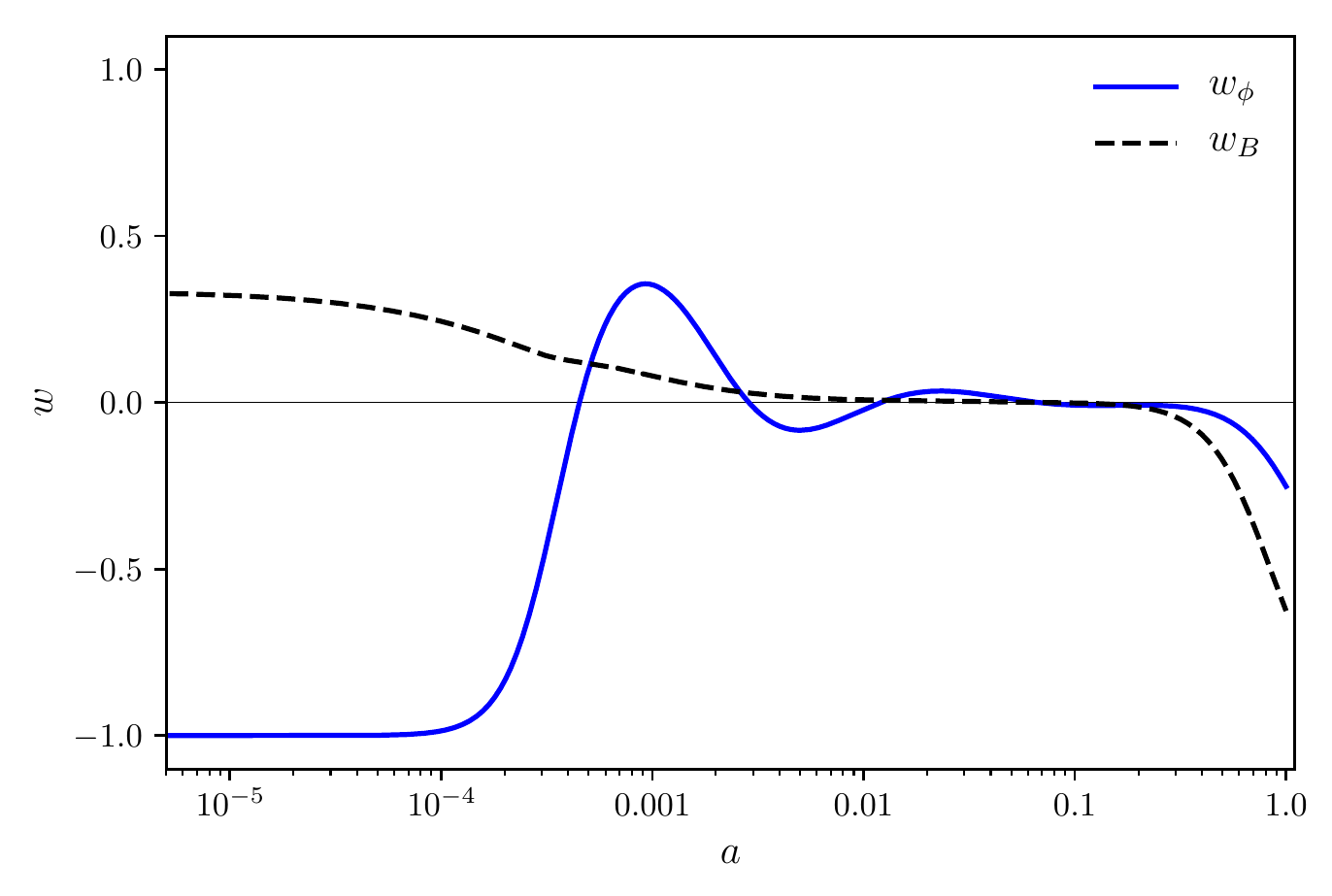}
\caption{\label{fig: weos} The evolution of the equation of state of the AQ field, $w_\phi$ (blue, solid), as a function of scale factor for a model with $\beta=12$ and $\mu=3.5$ Mpc$^{-1/2}$.  There is a downturn near $a\sim 0.4$ due to the onset of dark energy domination. For comparison, the background equation of state is shown (black, dashed).}
\end{figure}

The evolution of a single field in the exponential potential proceeds as follows. We consider the field to be initially frozen by the Hubble friction at $\phi=0$, in which case the equation of state is $w_\phi \approx -1$. The field begins to thaw and activate at a time determined by the parameter $\mu$ which is analogous to the $z_c$ of Ref.~\cite{Poulin2018ede}. The larger the value of $\mu$, the earlier it thaws. As the field evolves toward the attractor solution, the equation of state scales according to the dominant background component. In Fig.~\ref{fig: weos} we plot the evolution of $w_\phi$ as a function of scale factor for a field that becomes dynamical right around matter-radiation equality. As the field thaws, the equation of state jumps upwards to match the dominant component, initially overshooting its mark, before it settles to the matter-dominant evolutionary track.

The addition of multiple scaling fields in the AQ scenario changes the system dynamics \cite{Kim2005}. A succession of fields thaw and activate, each at a time determined by $\mu_i$. All active fields contribute to the energy density, each satisfying $\beta_i^2 \gg 1/3(1+w_B)$. However, the ensemble is characterized by an effective $\beta$,
\begin{equation}
    \frac{1}{\beta_{\rm eff}^2} = \sum_i \frac{1}{\beta_i^2}.
\end{equation}
As fields are successively thawed, $\beta_{\rm eff}$ is lowered, thereby raising the collective energy density. This continues until the bound on $\beta_{\rm eff}^2$ is saturated, when the fields ``flatten the potential" and inflate. At late times, the equation of state asymptotes to
\begin{equation}
    w_\phi = -1 + \frac{1}{3}\beta_{\rm eff}^2,
\end{equation}
approaching this limit from below \cite{Kim2005}.

Without the scaling behavior, the energy densities of the individual fields would be too small to ever dominate and acceleration would never arise.  In this way, AQ provides an ideal framework for EDE and dark energy. The necessary succession of thawing and scaling fields makes an early component plausible, and eventual cosmic acceleration inevitable.

There are many different ways to configure early and late dark energy components using $N$ fields, each introducing two parameters. In order to address the Hubble tension, we will consider a single early component that activates near matter-radiation equality. For simplicity, we will consider the remaining AQ fields to sufficiently resemble a component with $w_\phi\approx -1$ so that we may safely replace them with a cosmological constant.

\subsection{Resolving the Hubble Tension}

Early universe solutions to the Hubble tension are grounded in the theoretical description of the CMB. One of the best constrained features of the CMB anisotropy pattern is the angular size of the first acoustic peak, modeled as $\theta_s=r_s(z_*)/D_A(z_*)$. Here, $r_s(z_*)$ is the comoving sound horizon at decoupling, and $D_A(z_*)$ is the comoving angular diameter distance to the surface of last scattering, 
\begin{eqnarray}
\label{eq: sound horizon}
    r_s(z_*)&=&\int_{z_*}^\infty \frac{c_s dz'}{H(z')},\\
\label{eq: angular diameter}
    D_A(z)&=&\int_0^{z_*} \frac{dz'}{H(z')},
\end{eqnarray}
where $c_s$ is the sound speed. The sound horizon is dependent on pre-recombination energy densities and roughly scales with the Hubble parameter as $H_0^{-1/2}$, whereas $D_A(z_*)$ depends on densities after decoupling and scales as $H_0^{-1}$. This implies that if we decrease the sound horizon by adding new components to the energy density, and assuming the sound speed is unchanged, then we can increase the Hubble constant deduced from the CMB acoustic scale. As suggested in Ref.~\cite{Knox2019}, we focus on the brief window between matter-radiation equality and recombination as this is when the majority of the sound horizon accrues. By adding a small amount of EDE, it is possible to adequately lower the sound horizon, thereby increasing the Hubble constant inferred from the CMB.

\begin{figure}[t]
\includegraphics[width=0.47\textwidth]{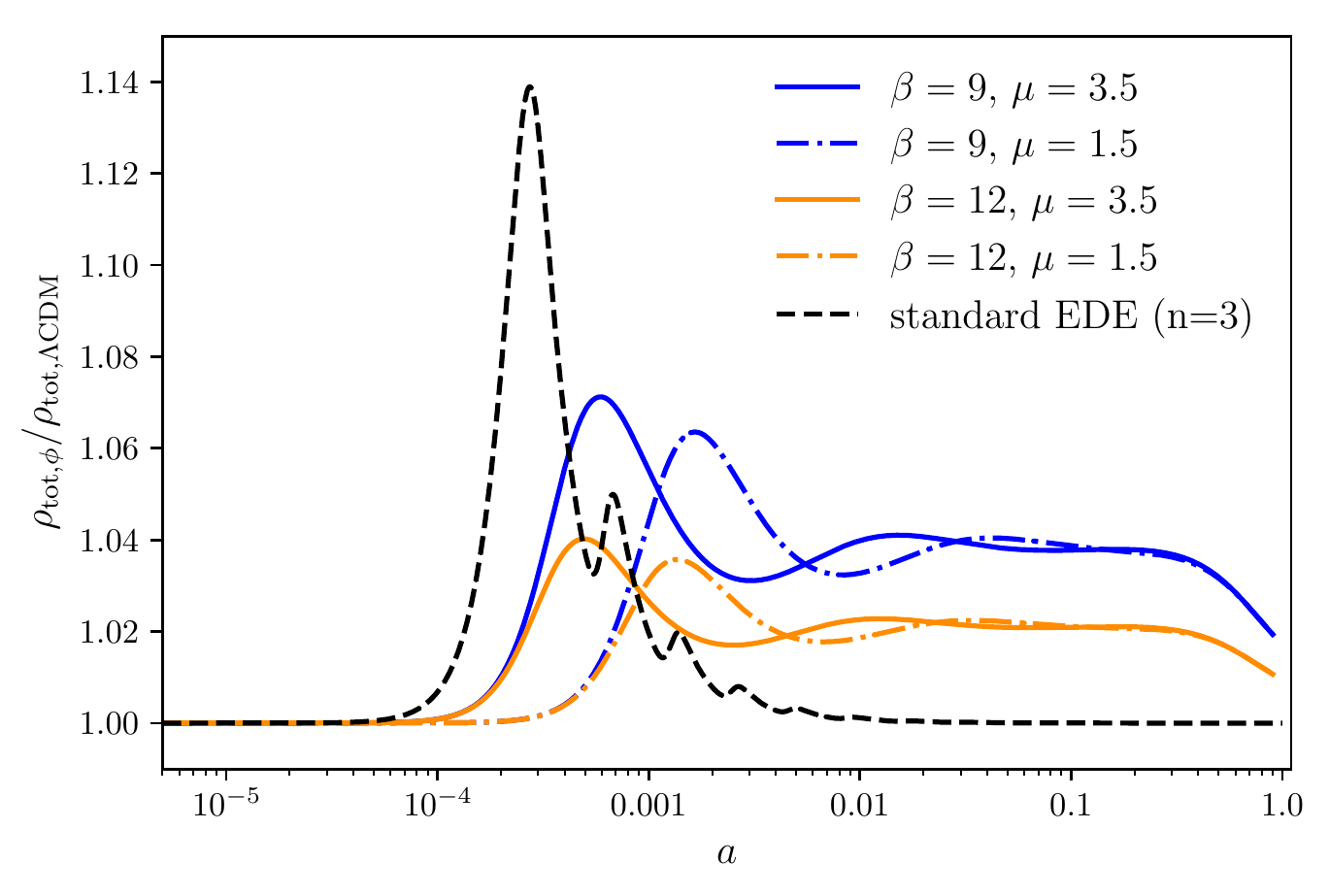}
\caption{\label{fig: background spike} The evolution of the fraction of the total energy density in the AQ model to the $\Lambda$CDM model as a function of scale factor. For higher $\beta$ and higher $\mu$, the contribution of the AQ field to the energy budget decreases and peaks earlier, respectively. Note that $\mu$ has units of Mpc$^{-1/2}$. The black-dashed line shows the best-fit n=3 oscillating scalar field model of EDE from Ref. \cite{Smith2019} for comparison. }
\end{figure}

An AQ field that activates during the epoch between equality and recombination, like the ones illustrated by the solid curves in Fig.~\ref{fig: background spike}, can produce enough of a spike in the total energy density to lower the sound horizon and raise the CMB inference of $H_0$ into agreement with local universe measurements. For example, based on Eqs.~(\ref{eq: sound horizon}-\ref{eq: angular diameter}), a model with $\beta = 12$ and $\mu = 3.5~{\rm Mpc}^{-1/2}$ and otherwise standard parameters should result in $H_0 \simeq 73$~km/s/Mpc. The use of a tracker potential has the added benefit of not demanding strict initial conditions, requiring only a two parameter extension to $\Lambda$CDM as opposed to the three parameter extensions required of other EDE models \cite{Karwal2016,Poulin2018ULA,Poulin2018ede,DEramo2018,Smith2019,Lin2019,Agrawal2019}. The overshoot in the equation of state helps sharpen the spike in energy, and afterwards the AQ field remains present at a trace level due to the matter-era scaling solution. 

\subsection{Linear Perturbations}

We have shown that at the background level an AQ field can resolve the Hubble tension. However, the viability of this scenario hinges on the behavior of the linear perturbations. For a single AQ field, the linear field perturbation $\delta\phi$ evolves according to the equation of motion
\begin{equation}
\label{eq: delta phi eom}
    \delta\phi'' +2\mathcal{H} \delta\phi'+ (k^2 +a^2V_{,\phi\phi})\delta\phi = -\frac{h'}{2}\phi',
\end{equation}
where $\mathcal{H}=a'/a$, the primes indicate the derivative with respect to conformal time $\prime = \partial/\partial\tau$, $h$ is the synchronous gauge metric potential (see \cite{Ma1995}), and we work in Fourier space. The system is equivalent to a damped, driven, harmonic oscillator. The homogeneous solution is negligible: any initial conditions set by inflation or other early universe processes have long been lost or erased as a consequence of the frozen field with $w_\phi \approx -1$ \cite{Dave:2002mn,Malquarti:2002bh}. Once the field begins to thaw, the inhomogeneous solution begins to take form, with an effective frequency of oscillation $\omega_\text{eff} = \sqrt{k^2+a^2V_{,\phi\phi}}$. 

To analyze the driving term, we focus on a field that thaws from the Hubble friction at or around matter-radiation equality so that the relevant evolution occurs in a matter-dominated background with $a \propto \tau^2$. We start from the well-known result that the CDM density contrast evolves in proportion to the scale factor, $\delta_c \propto a$, and that $h' = -2\delta_c'$ \cite{Ma1995}. From this, we infer that $h' \propto a' \propto \tau$. Next, according to the scaling solution in Eq.~(\ref{eqn:phi}), $\phi' = a \dot\phi \propto \tau^2/t$. Since conformal and cosmic time are related via $t\propto \tau^3$, we obtain $\phi' \propto \tau^{-1}$. Hence, the product $h' \phi'$ is independent of time. The driving term in Eq.~(\ref{eq: delta phi eom}) is constant as a result of the scaling solution for $\phi$.

\begin{figure}[t]
\includegraphics[width=0.47\textwidth]{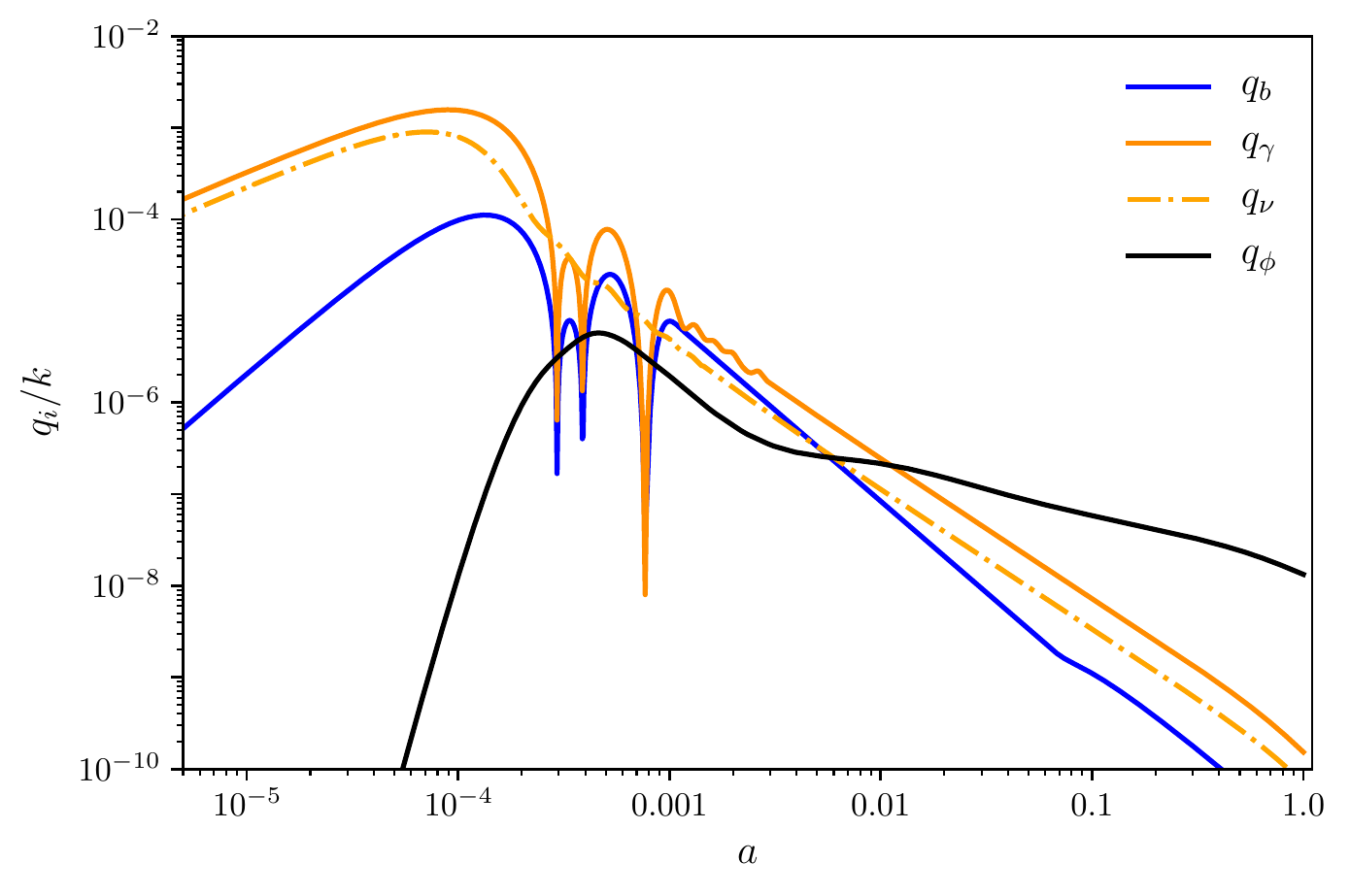}
\caption{\label{fig: dgq} The evolution of the heat flux of all relevant components as a function of scale factor for the $k=0.1$ Mpc$^{-1}$ wave mode. These curves are generated by a model with the best-fit parameter values taken from the ``semi-background'' AQ model run with $\theta_\phi$ allowed to evolve. The scaling behavior of the AQ field leads to the dominant contribution over the standard components, at $z \lesssim 100$ in the case shown.}
\end{figure}

There are two regimes of response to the constant driving term: for $\omega_\text{eff} \lesssim {\cal H}$, $\delta\phi$ grows in proportion to the scale factor; for $\omega_\text{eff} \gtrsim {\cal H}$, the perturbation solution is simply
\begin{equation}
    \delta\phi = -\frac{h' \phi'}{2(k^2 + a^2 V_{,\phi\phi})}.
\end{equation}
This solution divides into two cases. For the brief interval when $\omega_\text{eff} \gtrsim {\cal H}$ and $k^2 \ll a^2 V_{,\phi\phi}$, the scaling solution again dictates that $\delta\phi \propto a$, whereas at smaller scales, for $k^2 \gg a^2 V_{,\phi\phi}$, $\delta\phi$ is a constant. Hence, we have a simple story for the evolution of the AQ field perturbation: after an initial transient, $\delta\phi$ grows in proportion to the scale factor until the comoving mass scale drops below the wave number, $k^2 \gg a^2 V_{,\phi\phi}$, after which $\delta\phi$ is a constant. We note that if the AQ field decayed more rapidly than the background, then $\delta\phi$ would also decay. 

\begin{figure}[t]
\includegraphics[width=0.47\textwidth]{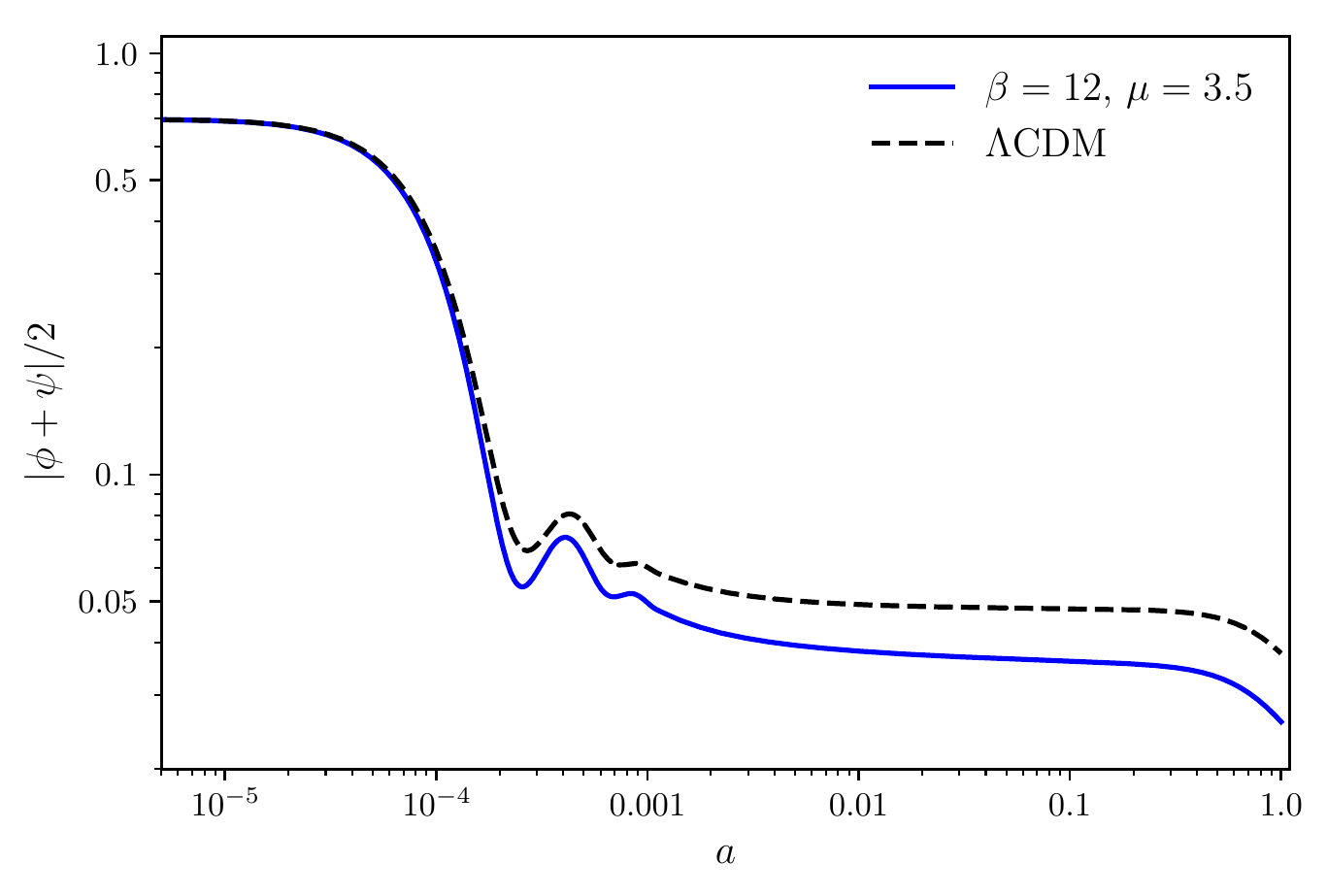}
\caption{\label{fig: weyl} Evolution of the Weyl gravitational potential as a function of scale factor for the $k=0.1$ Mpc$^{-1}$ wave mode. The solid blue curve shows the gravitational potential for a model with the best-fit parameters taken from the ``semi-background'' AQ model run with $\theta_\phi$ allowed to evolve. The best-fit $\Lambda$CDM model is shown by the black-dashed line. Due to the large heat flux of the scalar field, the gravitational potentials are shallower than in the $\Lambda$CDM model.  }
\end{figure}

\begin{table*}[t]
\begin{tabular}{| c | c | c | c |}
\hline \hline
 Parameter & $\Lambda$CDM & $\beta=12$, $\mu=3.5$ & $\beta=12$, $\mu=3.5$, $\theta_\phi=0$\\ \hline \hline
 $100 \omega_b$ & 2.235 (2.237) $\pm$ 0.015 & 2.182 (2.186) $\pm$ 0.014 & 2.068 (2.070) $\pm$ 0.016 \\ 
 $\omega_c$ & 0.1202 (0.1199) $\pm$ 0.0013  & 0.1297 (0.1294) $\pm$ 0.0013 & 0.1052 (0.1055) $\pm$ 0.0014 \\ 
 $100 \theta_s$ & 1.04089 (1.04105) $\pm$ 0.00032 & 1.04017 (1.04011) $\pm$ 0.00031 & 1.04220 (1.04212) $\pm$ 0.00033 \\
 $\tau$ & 0.0553 (0.0551) $\pm$ 0.0076 & 0.0594 (0.0596)$^{+0.0071}_{-0.0084}$ & 0.113 (0.106)$^{+0.015}_{-0.022}$\\ 
 $\ln (10^{10} A_s)$ & 3.046 (3.045) $\pm$ 0.015 & 3.069 (3.070)$^{+0.014}_{-0.016}$ & 3.121 (3.106)$^{+0.029}_{-0.040}$\\ 
 $n_s$ & 0.9645 (0.9644) $\pm$ 0.0043 & 0.9574 (0.9581) $\pm$ 0.0041 & 1.0010 (0.9994) $\pm$ 0.0053 \\
 $\beta$ & - & 12 (fixed) & 12 (fixed) \\
 $\mu$ [Mpc$^{-1/2}$] & - & 3.5 (fixed) & 3.5 (fixed) \\
 \hline
 $H_0$ [km/s/Mpc] & 67.27 (67.45) $\pm$ 0.56 & 64.20 (64.29) $\pm$ 0.55 & 73.12 (72.98) $\pm$ 0.77 \\ 
 $S_8$ & 0.834 (0.829) $\pm$ 0.013 & 0.858 (0.856) $\pm$ 0.013 & 0.726 (0.723) $\pm$ 0.013 \\
 \hline
 Total $\chi^2_\text{min}$ & 1014.09 & 1048.38 & 1307.81\\
 \hline \hline 
\end{tabular}
\caption{\label{tab: P18 constraints} The mean (best-fit) $\pm 1\sigma$ error of the cosmological parameters for $\Lambda$CDM, the AQ model with $\beta=12$, $\mu=3.5$ Mpc$^{-1/2}$ and the ``semi-background'' model with $\theta_\phi=0$. Constraints are based on the full \textit{Planck} 2018 dataset. }
\end{table*}

We can use the results of this simple analysis to forecast the behavior of the AQ field perturbations in terms of fluid variables. The most significant role is played by the heat flux $q_\phi=8 \pi G a^2(\rho_\phi + p_\phi)\theta_\phi$, where $\theta_\phi$ is the velocity divergence of the AQ field. The heat flux obeys the equation 
\begin{equation}
\label{eq: heat flux}
    q_\phi' + 2 \mathcal{H}q_\phi = 8 \pi G a^2 k^2\delta p_\phi, 
\end{equation}
where $\delta p_\phi$ is the pressure perturbation. The heat flux is related to our AQ field via $\theta_\phi$ through  
\begin{equation}
    (\rho_\phi+p_\phi)\theta_\phi = \frac{k^2}{a^2} \phi' \delta\phi.
\end{equation}
Again, we can use the scaling solution $\phi' \propto a^{-1/2}$ and $\delta\phi \propto a$ , in which case the heat flux grows $\propto a^{1/2}$, until the comoving mass scale drops below the wave number. Thereafter $q_\phi$ decays $\propto a^{-1/2}$. This is significant, because all other contributions due to CDM, baryons, photons, neutrinos are zero (CDM) or decay more rapidly, and will eventually grow subdominant to the scalar field contribution. An example based on our numerical calculations is shown in Fig.~\ref{fig: dgq}. Despite contributing to the energy budget at a percent level, the AQ field has an outsize effect. 

On the same scales, the AQ density perturbation $\delta \rho_\phi$ loses energy, decaying at the same rate as the background so that $\delta_\phi=\delta\rho_\phi/\rho_\phi$ is constant. Moreover, the pressure perturbation is $\delta p_\phi \approx - \delta \rho_\phi$, like a tension. Hence, the fluctuation response of the scalar field inhibits clustering.

The AQ contribution to the heat flux sources the trace-free scalar metric perturbation, $\eta$ \cite{Ma1995}. In more physical terms, it causes the post-recombination gravitational potentials to decay, resulting in an additional integrated Sachs-Wolfe effect, an example of which is shown in Fig.~\ref{fig: weyl}. Due to the timing of this behavior, it primarily affects modes that determine the shape of the CMB anisotropy pattern at degree scales and larger. But there are more facets to the ultimate impact on the predicted CMB temperature and polarization ansiotropy pattern, which we turn to next.

\section{Data and Methodology}
\label{sec: data and methods}

We run a complete Markov Chain Monte Carlo (MCMC) using the public code \texttt{CosmoMC} (see \url{https://cosmologist.info/cosmomc/}) \cite{Lewis2002} interfaced with a modified version of \texttt{CAMB} to directly solve the linearized scalar field equations \cite{Lewis1999}. Details are provided in Appendix~\ref{sec: num implement}. We model the neutrinos as two massless and one massive species with $m_\nu$ = 0.06 eV and $N_\text{eff} = 3.046$. We use a dataset consisting of \textit{Planck} 2018 measurements of the CMB via the \texttt{TTTEEE Plik lite} high-$\ell$, \texttt{TT} and \texttt{EE} low-$\ell$, and lensing likelihoods \cite{Aghanim2019}.
The \texttt{Plik lite} likelihood is a foreground and nuisance marginalized version of the \texttt{Plik} likelihood \cite{Aghanim2019}. We have found that the two datasets return nearly identical posterior distributions for a typical AQ model. Based on this, we infer that the AQ model has negligible effect on the \textit{Planck} nuisance parameters, allowing us to use the \texttt{lite} likelihood in place of the full likelihood, and speeding up our MCMC analysis.
We restrict ourselves to only CMB data to determine whether a scaling field can independently raise the CMB-derived value of the Hubble constant, without the influence of late universe measurements, although we give results for extended datasets in Appendix~\ref{sec: extended results fixed beta mu}.  

We perform an analysis with a Metropolis-Hastings algorithm with flat priors on the six standard cosmological parameters $\{ \omega_b, \omega_c, \theta_s, \tau, \ln(10^{10} A_s), n_s \}$ as well as the model parameters $\beta$ and $\mu$. Our results are obtained by running eight chains and monitoring convergence via the Gelman-Rubin criterion, with $R-1<0.05$, for all parameters, being considered complete convergence \cite{GelmanRubin1992}. Throughout this paper we absorb a factor of $(8\pi G)^{1/2}$ into the parameter $\beta$, allowing us to report it as a unitless scale parameter, matching its implementation within \texttt{CAMB}. Similarly, we report $\mu$ in units of Mpc$^{-1/2}$ where we absorb a factor of $(8\pi G)^{1/4}$.

\section{Observational Constraints}
\label{sec: observational constraints}

In this section we explore the implications of adding an AQ field for CMB-derived cosmological parameters. For fixed $\beta$ and $\mu$ we show that the homogeneous AQ field can provide a resolution to the Hubble tension. However, the scaling behavior leads to strong perturbations that spoil the concordance. We then explore the model parameter space and show that the data ultimately prefers AQ models that resemble $\Lambda$CDM.

\subsection{Fixed Model Parameters}
\label{sec: fixed beta and mu}

\begin{figure*}[t]
\includegraphics[width=0.96\textwidth]{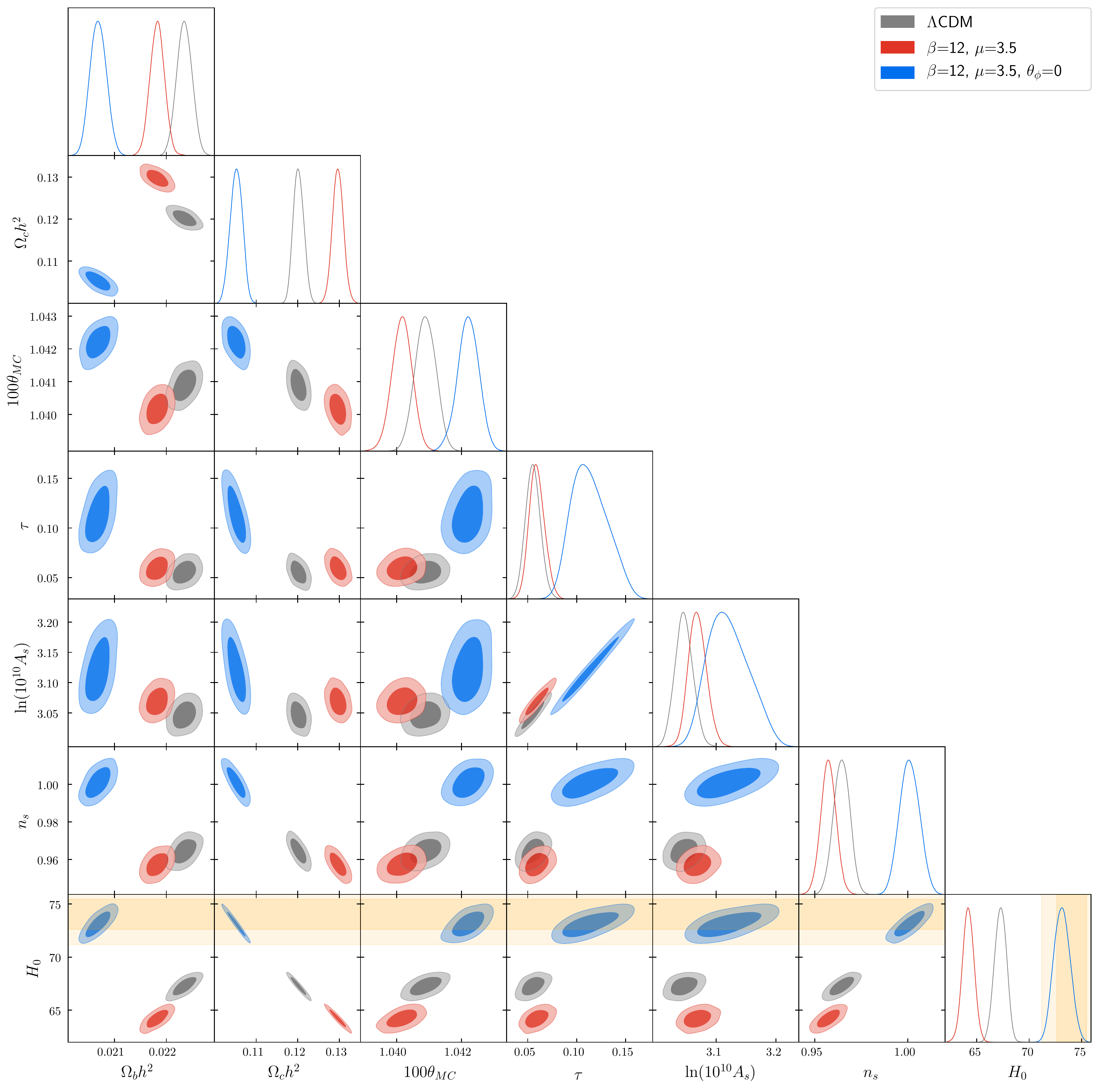}
\caption{\label{fig: tri-plot} Posterior distributions of the AQ model with $\beta=12$, $\mu=3.5$ Mpc$^{-1/2}$ and $\theta_\phi$ turned on (red) and off (blue), and the $\Lambda$CDM model (gray) for the \textit{Planck} 2018 dataset. The darker inner (lighter outer) regions correspond to 1$\sigma$ (2$\sigma$) confidence intervals. The SH0ES determination of $H_0$ is shown in the orange bands.}
\end{figure*}

\begin{figure*}[t]
\includegraphics[width=0.66\textwidth]{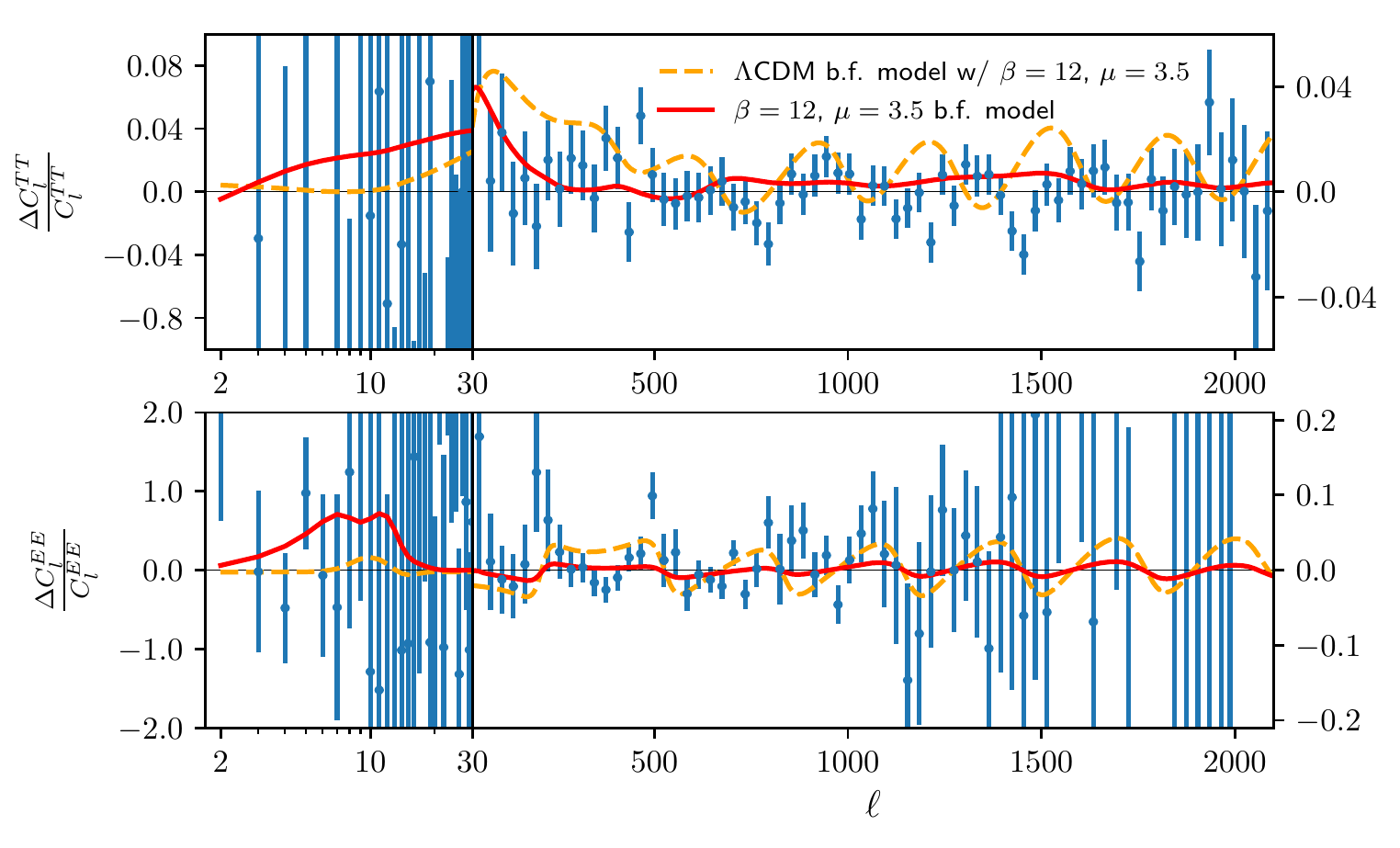}
\caption{\label{fig: residuals} Temperature and polarization power spectrum residuals between the best-fit $\Lambda$CDM model and the best-fit AQ cosmology (solid red), as well as an AQ cosmology with the six standard model parameters unchanged from their best-fit $\Lambda$CDM values (dashed orange). We show the residuals from \textit{Planck} 2018 data in blue. Left (right) vertical axis scaling is for multipoles less (greater) than 30.}
\end{figure*}

We fix the model parameters to $\beta=12$ and $\mu=3.5$~Mpc$^{-1/2}$ such that the AQ field provides an approximately $4$\% spike in the background energy density in the epoch between matter-radiation equality and recombination, as shown in Fig.~\ref{fig: background spike}. An early contribution of this size should be enough to raise the value of the Hubble constant inferred by CMB measurements \cite{Knox2019,Poulin2018ede}. 

We consider two alternative models for comparison. The first is $\Lambda$CDM, as a control. The second is also an AQ model with $\beta=12$ and $\mu=3.5$~Mpc$^{-1/2}$, but for which the AQ velocity divergence is artificially set to zero, $\theta_\phi=0$. We refer to this model as ``semi-background". Without the inclusion of the velocity divergence, this model is self-inconsistent. However, we find the model to be helpful to illustrate the influence of the velocity divergence on cosmological parameters in this scenario. Note that the inclusion of the density perturbation of the field has little effect on the temperature and polarization anisotropies since the total energy density perturbation is dominated by CDM.

The results of the MCMC analysis, consisting of constraints to the cosmological parameters for the AQ cosmology, the ``semi-background" AQ cosmology, and $\Lambda$CDM, are presented in Table \ref{tab: P18 constraints}. We show the posterior distributions for the relevant parameters in these models in Fig.~\ref{fig: tri-plot}. 

These constraints paint an interesting picture. The AQ ``semi-background" model yields a best-fit value of $H_0=72.98$ km/s/Mpc, in excellent agreement with the SH0ES determination of $H_0$. Hence, our initial rationale for selecting this model is justified. However, the quality of the fit to the data is poor compared to $\Lambda$CDM, as seen in the increased $\chi^2_\text{min}$. This is nearly entirely due to the self-inconsistency of the model. Without the complete evolution of field perturbations, terms that normally cancel the strong, late-time ISW effect in the CMB are absent leading to a huge increase in power in the large scale CMB anisotropy pattern \cite{Caldwell:2000wt}. Restoring the velocity divergence, the AQ cosmology with $\beta=12$ and $\mu=3.5$ Mpc$^{-1/2}$ yields a surprise. The model not only fails to solve the Hubble tension but exacerbates it even further, giving a best-fit value of $H_0=64.29$ km/s/Mpc as shown in Table \ref{tab: P18 constraints}. What these results suggest is that at the homogeneous level, the spike in the energy density given by the AQ field would indeed raise the CMB inferred value of the Hubble constant. But the dominant role of the AQ contribution to the heat flux spoils the concordance.

We can now take a sharper look at the role of the AQ perturbations, with the benefit of hindsight of the parameter analysis. We use the best-fit parameter values determined for the ``semi-background" model, and apply them to the full AQ model. This enables us to see the effect of the heat flux on the metric perturbations and the full CMB anisotropy.

During the matter era, the density contrast of the AQ field is constant, meaning the density perturbation $\delta\rho_\phi$ must be losing energy. This is matched by the growth of the heat flux  $q_\phi=8\pi G a^2(\rho_\phi+p_\phi)\theta_\phi$, shown in Fig.~\ref{fig: dgq}. This behavior has the same effect as the free-streaming of photons and neutrinos out of potential wells, bringing energy with them as they go. This outflow of energy causes rarefaction of the gravitational potential wells when compared to $\Lambda$CDM, as shown in Fig.~\ref{fig: weyl}. 

The change in the potential wells has widespread consequences. Most importantly for the Hubble tension, there is now an additional integrated Sachs-Wolfe (ISW) effect driven by the AQ heat flux. For the parameters of these models, this new ISW raises the power of the CMB spectrum across the first acoustic peak. To compensate for this change, there is a series of parameter changes when compared to $\Lambda$CDM, as shown in Table \ref{tab: P18 constraints}. Most notably, the CDM density is increased, which introduces a phase shift in the acoustic oscillations toward larger angular scales for all multipoles. To maintain the correct angular scale of the acoustic peaks, $H_0$ is lowered. 

The residual between the best-fit $\Lambda$CDM model and our AQ model, shown in Fig.~\ref{fig: residuals}, makes these parameter changes clearer. In the solid red line we show the residual for the best-fit AQ model using the full \textit{Planck} dataset and in orange we show the residual for the AQ model using standard model parameters specified by the best-fit $\Lambda$CDM model. Setting the standard model parameters to their $\Lambda$CDM values and adding in an AQ field allows us to illustrate the full influence of the AQ field on the CMB spectrum. In the AQ model with $\Lambda$CDM parameters, the oscillation in the residual seen at high-$\ell$ in both temperature and polarization shows a phase-shift toward high-$\ell$, which can be remedied by a higher value of $H_0$. However, the additional ISW effect caused by the domination of the heat flux of the AQ field, seen most clearly between $10<\ell<400$, is too strong to overcome. When we shift the parameter values to match the best-fit AQ model, the higher CDM density and lower value of $H_0$ deepen the gravitational potentials, and restore the angular scale of acoustic oscillations, resulting in a closer match to the data, although one that is still not on par with $\Lambda$CDM. This poor fit is consistent among the individual likelihoods in the \textit{Planck} 2018 dataset (to remind, these are: high-$\ell$ TT,TE,EE; low-$\ell$ TT; low-$\ell$ EE; lensing). The biggest deviation comes from the high-$\ell$ TT,TE,EE likelihood with $\Delta \chi^2_{min} \approx 25$, supported by the offset of the best-fit AQ model residuals from the \textit{Planck} 2018 data points in Fig.~\ref{fig: residuals}. 

The overall change in level of the gravitational potentials also leaves imprints on the matter power spectrum, which can be summarized through the parameter $S_8=\sigma_8 (\Omega_m/0.3)^{1/2}$. Weak lensing surveys like KiDS-450 measure $S_8=0.745\pm 0.039$ \cite{Hildebrandt:2016iqg}. This is in a $\sim 2 \sigma$ tension with the high value of $S_8=0.832\pm 0.013$ estimated by \textit{Planck} using $\Lambda$CDM \cite{Aghanim2018}. We can see from Table \ref{tab: P18 constraints} that the ``semi-background'' model lowers the value of $S_8$ estimated by \textit{Planck} data into agreement with the local universe measurement from KiDS-450. However, similarly to $H_0$, when the velocity divergence is restored to the AQ model, this concordance is lost, and the tension is exacerbated.

With a lower preferred value of $H_0$ and a worse fit to the full \textit{Planck} 2018 dataset, it seems that the scaling behavior present in this model, which provides a natural link between early and late dark energy as well as a framework to solve the ``why now'' problem, is the very mechanism that spoils this model as a solution to the Hubble tension.

\subsection{Full Results}

\begin{figure*}[t]
\includegraphics[width=\textwidth]{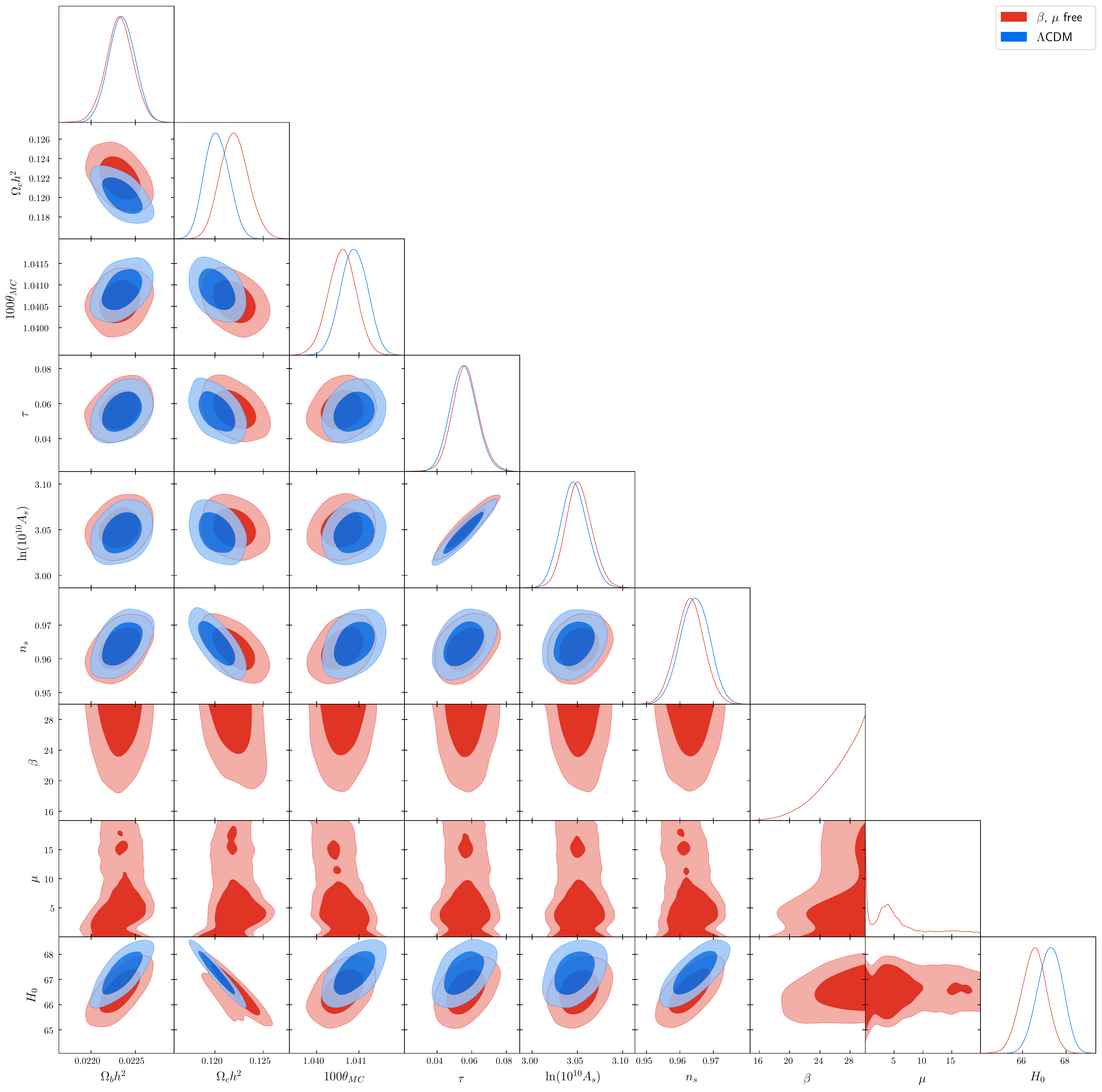}
\caption{\label{fig: full8 tri-plot} Posterior distributions of the AQ model with $\beta$, and $\mu$ as free parameters (red) and the $\Lambda$CDM model (blue) for the \textit{Planck} 2018 dataset. The darker inner (lighter outer) regions correspond to 1$\sigma$ (2$\sigma$) confidence intervals.}
\end{figure*}

We now promote $\beta$ and $\mu$ to free parameters and allow them to vary alongside the six standard model parameters with flat priors, $5<\beta<30$ and $0.0001<\mu<20$ Mpc$^{-1/2}$. We leave out $\mu=0$ for numerical stability within \texttt{CAMB}. The parameter constraints derived from the \textit{Planck} 2018 dataset are presented in Table \ref{tab: full8 P18 constraints}, with posterior distributions for the relevant parameters shown in Fig.~\ref{fig: full8 tri-plot}. We include the posteriors for the best-fit $\Lambda$CDM model for comparison.  

It is clear from Fig.~\ref{fig: full8 tri-plot} that the data prefers high values of $\beta$ and low values of $\mu$. For the best-fit value of $\beta=23.1$, the AQ field constitutes $<1\%$ of the total energy density once it settles to its scaling solution. For such a low density component, the time that the AQ field thaws from the Hubble friction is inconsequential, resulting in a wide spread in the posterior distribution of $\mu$. However, the best-fit value of $\mu=0.005$ Mpc$^{-1/2}$ and the preference for $\mu<7.07$ Mpc$^{-1/2}$ gives us some insight into these results.  

As previously discussed, low values of $\mu$ correspond to later activation of the AQ field, meaning the field behaves like a cosmological constant with a negligible energy density for most of its evolution. For the best-fit values of $\beta=23.1$ and $\mu=0.005$ Mpc$^{-1/2}$, the AQ field thaws from the Hubble friction during dark energy domination at which time its scaling behavior forces it to behave as an additional, subdominant cosmological constant. In this case, the field forgoes the post-recombination decay of gravitational potentials caused by the domination of the heat flux of the AQ field during the matter-era, allowing for the best-fit matter densities to remain unchanged from their values in the $\Lambda$CDM model. However, Fig.~\ref{fig: full8 tri-plot} shows us that even an AQ field present with a small abundance shifts the peak of the posterior distribution of $H_0$ toward smaller values, furthering the evidence that this model cannot resolve the Hubble tension.

The best-fit AQ cosmology, introducing a new component that makes up $<1\%$ of the total energy density, is statistically indistinguishable from $\Lambda$CDM with $\Delta \chi^2_\text{min}=-0.53$. These results tell us there is little to no evidence for the presence of an AQ scaling field within the full \textit{Planck} 2018 dataset. 
\begin{table}[t]
\begin{tabular}{| c | c |}
\hline \hline
 Parameter              & $\beta$, $\mu$ free \\ \hline \hline
 $100 \omega_b$         & 2.232 (2.237) $\pm$ 0.015 \\ 
 $\omega_c$             & 0.1220 (0.1199)$^{+0.0014}_{-0.0015}$ \\ 
 $100 \theta_s$         & 1.04061 (1.04078) $\pm$ 0.00033 \\
 $\tau$                 & 0.0565 (0.0543) $\pm$ 0.0076 \\ 
 $\ln (10^{10} A_s)$    & 3.052 (3.049) $\pm$ 0.015 \\ 
 $n_s$                  & 0.9630 (0.9647) $\pm$ 0.0042 \\
 $\beta$                & $>25.6$ (23.1) \\
 $\mu$ [Mpc$^{-1/2}$]   & $<7.07$ (0.005) \\
 \hline
 $H_0$ [km/s/Mpc]       & 66.52 (67.22) $\pm$ 0.58 \\ 
 $S_8$                  & 0.839 (0.833) $\pm$ 0.013 \\
 \hline
 Total $\chi^2_\text{min}$ &  1013.56 \\
 \hline \hline 
\end{tabular}
\caption{\label{tab: full8 P18 constraints} The mean (best-fit) $\pm 1\sigma$ error of the cosmological parameters in the full AQ model analyzed using the \textit{Planck} 2018 dataset. }
\end{table}

We note that the changes to the gravitational potentials discussed in Sec.~\ref{sec: fixed beta and mu} also affect the imprint of gravitational lensing on the CMB power spectrum. As CMB photons travel along the line-of-sight, they are gravitationally deflected by the large-scale distribution of matter in the Universe. This lensing effect blurs the anisotropy pattern and smooths the acoustic peaks. When we artificially turn off the effects of CMB lensing and use only \textit{Planck} high-$\ell$ TT, TE, EE, and low-$\ell$ TT and EE data, we find that the AQ model provides a statistically better fit to the data than $\Lambda$CDM. Results for this analysis in the AQ model with free $\beta$ and $\mu$ are shown in Appendix \ref{sec: extended results free beta mu}. However, due to the scaling of the AQ field, the gravitational potentials are shallower, implying less blurring and smoothing. Turning lensing back on results in a poorer relative fit to data than $\Lambda$CDM. When considering the $S8$ tension between \textit{Planck} and local universe weak lensing surveys, and the $A_L$ anomaly present in \textit{Planck} data \cite{Motloch:2019gux}, these results become more interesting and may warrant further investigation.

\section{Discussion and Conclusions}
\label{sec: conclusion}

The Hubble tension has motivated a variety of extensions to the standard $\Lambda$CDM cosmological model, most of which focus on injecting energy at or around the time of matter-radiation equality. In this paper we considered the possibility that a scaling field which activates just prior to recombination provides this energy injection. Specifically, we evaluated the impact on the CMB of a scalar field with an exponential tracking potential of the form $V(\phi)=\mu^4 e^{-\beta\phi}$ in the context of an AQ scenario for EDE and DE. This model constitutes a two parameter extension to the standard $\Lambda$CDM model specified by the steepness of the potential $\beta$, and the effective mass of the field $\mu$. In this scenario, early dark energy is simply a sign of the build up of dark energy.

The Hubble tension would appear ameliorated at the background level by a scenario with $\beta = 12$ and $\mu = 3.5$ Mpc$^{-1/2}$. Solving for the dynamics of the linearized perturbations of the field, however, we find a different story. The scaling behavior of the AQ field results in the domination of the heat flux of the AQ field over that of the standard model components. The impact on the CMB power spectrum actually worsens the Hubble tension to an almost $7 \sigma$ difference with local universe measurements. Ultimately, we find that \textit{Planck} 2018 temperature and polarization data, plus \textit{Planck} estimates of the lensing potential, constrain the AQ model parameters to resemble a $\Lambda$CDM-like cosmology; the best-fit AQ model is statistically indistinguishable from $\Lambda$CDM. 
 
The failure of this model offers insight into the ability of new physics to resolve the Hubble tension.  In this case, the pressure fluctuation drives the growth of the heat flux on subhorizon scales as shown in Eq.~\ref{eq: heat flux}. This sets off a cascade of effects, softening the gravitational potentials, shifting the acoustic peaks in the CMB, and ultimately exacerbating the Hubble tension. A few ways around this result are suggested. For example, if we abandon the scaling solution and use a model that spikes just prior to recombination and then decays faster than the background, then the pressure source decays, too. This is the method employed in Refs.~\cite{Poulin2018ede,Agrawal2019,Smith2019,Braglia:2020bym}. The price of which is an additional parameter, which may require the fine tuning of the initial conditions. Another solution would be to introduce an additional term on the right hand side of Eq.~\ref{eq: heat flux} to damp or diminish the pressure. This might be accomplished by coupling to another field \cite{DiValentino2017c}. Yet neither of these fixes do more than soften the Hubble tension.

One would expect that a compelling solution would raise the CMB-inferred value of $H_0$ into complete agreement with local universe measurements while also addressing (or at the very least, not exacerbating) other cosmological tensions. This may require dropping the scalar field as a possible solution. If EDE does play a role in resolving the various tensions between $\Lambda$CDM and cosmological observations, it will necessarily have more structure than the most basic scenarios that have been considered. Currently, no model has succeeded at independently and adequately solving the Hubble tension. Improved measurements of the CMB, H(z), and BAO at various redshifts will give us better insight into possible physics beyond the standard model.

\begin{acknowledgments}
We thank Jose Luis Bernal, Evan McDonough, and Tristan Smith for useful comments. This work is supported in part by U.S. Department of Energy Award No. DE-SC0010386. 

\end{acknowledgments}

\appendix

\section{Numerical Implementation}
\label{sec: num implement}

In \texttt{CAMB}, the evolution of the perturbation equations and the construction of the angular power spectra require the background densities of all components to be specified throughout cosmic history. The standard model background densities all follow simple scaling relations for which only the present day density is needed to completely specify their evolution. The background evolution of the AQ field is non-trivial and must be numerically solved prior to the evolution of the perturbations in order to obtain accurate results. The background evolution of the field is specified by the homogeneous Klein-Gordon equation, which requires initial conditions on $\phi$ and $\phi'$ in order to be evolved. For the exponential potential we can absorb the initial value of $\phi$ into the parameter $\mu$, allowing us to set $\phi_i=0$. The use of a tracker potential means that for a wide range of initial values of $\phi'$, the field will settle to its attractor solution, hence we can arbitrarily set $\phi'=0$, following slow-roll conditions. With initial conditions set, we numerically solve Eq. (\ref{eq: homogeneous KG}) to create arrays of values for $\phi$ and $\phi'$ over cosmic time which we interpolate whenever background values for the field are needed during the evolution of the cosmological perturbations. 

The evolution of the AQ field fluctuations are solved numerically alongside the standard model perturbations. To properly interface the scalar field with the standard model components, we need to translate the field perturbations into fluid variables. Linearly perturbing the scalar field stress-energy tensor yields:
\begin{eqnarray}
\label{eq: deltarho}
    \delta\rho_\phi &=& a^{-2}\phi'\delta\phi'+V_{,\phi}\delta\phi, \\
    \delta p_\phi &=& a^{-2}\phi'\delta\phi'-V_{,\phi}\delta\phi,\\
    (\rho_\phi+p_\phi)\theta_\phi &=& \frac{k^2}{a^2} \phi' \delta\phi,\\
\label{eq: sigma}
    (\rho_\phi+p_\phi)\sigma_\phi &=& 0,
\end{eqnarray}
where $\theta_\phi=i k^j v_j$ represents the velocity divergence, and $\sigma_\phi$ is the anisotropic stress, which is zero for a scalar field. Using these, we can show that the conservation of the scalar field stress-energy follows that of a single uncoupled fluid \cite{Ma1995}:
\begin{eqnarray}
    \delta_\phi' &=& -(1+w_\phi)\left(\theta_\phi+\frac{h'}{2}\right)-3\mathcal{H}\left(\frac{\delta p_\phi}{\delta\rho_\phi}-w_\phi\right)\delta_\phi, \quad \\
    \theta_\phi' &=& -\mathcal{H}(1-3w_\phi)\theta_\phi - \frac{w_\phi'}{1+w_\phi}\theta_\phi + \frac{\delta p_\phi/\delta\rho_\phi}{1+w_\phi} k^2 \delta_\phi, \qquad 
\end{eqnarray}
where $\delta_\phi=\delta\rho_\phi/\rho_\phi$ and $\delta p_\phi/\delta\rho_\phi = c_\phi^2$ gives the adiabatic sound speed squared. While these fluid equations of motion are mathematically equivalent to the linearized KG equation, they are numerically unstable in practice since $w_\phi = -1$ prior to the slow roll of the field. In our numerical implementation we instead directly evolve the linearized KG equation, Eqs.~(\ref{eq: delta phi eom}), and construct the necessary fluid variables using Eq. \ref{eq: deltarho}-\ref{eq: sigma}.

\begin{table*}[t]
\begin{tabular}{| c | c | c |}
\hline \hline
 CMB+BAO & $\Lambda$CDM & $\beta=12$, $\mu=3.5$ \\ \hline \hline
 $100 \omega_b$ & 2.240 (2.240) $\pm$ 0.013 & 2.217 (2.220) $\pm$ 0.014  \\ 
 $\omega_c$ & 0.11947 (0.11954) $\pm$ 0.00097   & 0.1247 (0.1246) $\pm$ 0.0011 \\ 
 $100 \theta_s$ & 1.04098 (1.04103) $\pm$ 0.00030 & 1.04075 (1.04086) $\pm$ 0.00028 \\
 $\tau$ & 0.0573 (0.0570) $\pm$ 0.0074 & 0.0739 (0.0736) $\pm$ 0.0093 \\ 
 $\ln (10^{10} A_s)$ & 3.049 (3.048) $\pm$ 0.014 & 3.089 (3.090) $\pm$ 0.018 \\ 
 $n_s$ & 0.9663 (0.9659) $\pm$ 0.0038 & 0.9685 (0.9688) $\pm$ 0.0038 \\
 $\beta$ & - & 12 (fixed) \\
 $\mu$ [Mpc$^{-1/2}$] & - & 3.5 (fixed) \\
 \hline
 $H_0$ [km/s/Mpc] & 67.59 (67.58) $\pm$ 0.44 & 66.33 (66.43) $\pm$ 0.46  \\ 
 $S_8$ & 0.828 (0.828) $\pm$ 0.011 & 0.815 (0.814) $\pm$ 0.011 \\
 \hline
 Total $\chi^2_\text{min}$ & 1020.02 & 1082.89 \\
 \hline \hline 
\end{tabular}
\caption{\label{tab: CMB+BAO} The mean (best-fit) $\pm 1\sigma$ error of the cosmological parameters for $\Lambda$CDM and the AQ model with $\beta=12$, $\mu=3.5$ Mpc$^{-1/2}$. Constraints are based on the CMB and BAO datasets. }
\end{table*}

\begin{table*}[t]
\begin{tabular}{| c | c | c |}
\hline \hline
 CMB+BAO+R19 & $\Lambda$CDM & $\beta=12$, $\mu=3.5$ \\ \hline \hline
 $100 \omega_b$ & 2.252 (2.250) $\pm$ 0.013 & 2.231 (2.230) $\pm$ 0.014  \\ 
 $\omega_c$ & 0.11837 (0.11819) $\pm$ 0.00095 & 0.12317 (0.12313) $\pm$ 0.00098 \\ 
 $100 \theta_s$ & 1.04113 (1.04117) $\pm$ 0.00029 & 1.04092 (1.04105) $\pm$ 0.00028 \\
 $\tau$ & 0.0608 (0.0589) $^{+0.0071}_{-0.0080}$ & 0.081 (0.079) $\pm$ 0.010 \\ 
 $\ln (10^{10} A_s)$ & 3.054 (3.051) $^{+0.014}_{-0.016}$ & 3.100 (3.095) $\pm$ 0.019 \\ 
 $n_s$ & 0.9689 (0.9702) $\pm$ 0.0037 & 0.9721 (0.9728) $\pm$ 0.0039 \\
 $\beta$ & - & 12 (fixed) \\
 $\mu$ [Mpc$^{-1/2}$] & - & 3.5 (fixed) \\
 \hline
 $H_0$ [km/s/Mpc] & 68.13 (68.18) $^{+0.39}_{-0.43}$ & 67.05 (67.09) $\pm$ 0.43  \\ 
 $S_8$ & 0.817 (0.814) $\pm$ 0.011 & 0.803 (0.801) $\pm$ 0.011 \\
 \hline
 Total $\chi^2_\text{min}$ & 1039.15 & 1109.79 \\
 \hline \hline 
\end{tabular}
\caption{\label{tab: CMB+BAO+R19} The mean (best-fit) $\pm 1\sigma$ error of the cosmological parameters for $\Lambda$CDM and the AQ model with $\beta=12$, $\mu=3.5$ Mpc$^{-1/2}$. Constraints are based on the CMB, BAO, and R19 datasets. }
\end{table*}

In \texttt{CAMB}, distance and time are measured in Mpc. For numerical simplicity, we absorb a factor of $(8\pi G)^{1/2}$ into $\beta$ and a factor of $(8\pi G)^{1/4}$ into $\mu$ so that 
\begin{equation}
    \frac{1}{2}\left(\frac{\phi'}{a^2}\right)^2+V(\phi) = 8 \pi G \rho
\end{equation}
has units of Mpc$^{-2}$. To convert to standard particle physics units, remember that 1 Mpc $=1.5637\times 10^{38}$  GeV$^{-1}$. Converting the model parameters presented in Sec.~\ref{sec: fixed beta and mu} from \texttt{CAMB} units to particle physics units gives $\beta = 5 \times 10^{-18}$/GeV, and $\mu = 0.43$ eV.

\section{Extended Results for Fixed Model Parameters}
\label{sec: extended results fixed beta mu}

In this Appendix we present an extended MCMC analysis on the AQ model with fixed model parameters. To fully analyze the cosmological impact of the AQ model we use a wider range of cosmological datasets for parameter estimation:
\begin{itemize}
    \item {\bf Cosmic Microwave Background (CMB):} We use the Planck 2018 measurements of the CMB (via \texttt{TTTEEE Plik lite} high-$\ell$, \text{TT} and \text{EE} low-$\ell$, and lensing likelihoods \cite{Aghanim2019},
    \item {\bf Baryon acoustic oscillation (BAO) data:} We use data from the BOSS survey (data release 12) at $z=0.38,0.51,$ and $0.61$ \cite{Alam2016}, low redshift measurements from the 6dF survey at $z=0.106$ \cite{Beutler2011}, and the BOSS main galaxy sample at $z=0.15$ \cite{Ross2014},
    \item {\bf Local Hubble constant measurement (R19):} The measurement of the local Hubble constant giving $H_0=74.03\pm1.42$ km/s/Mpc the SH0ES collaboration \cite{Riess2019}. 
\end{itemize}

\begin{table}[b]
\begin{tabular}{ | c | c | c | }
\hline \hline
Dataset & $\Lambda$CDM & $\beta=12$, $\mu=3.5$ \\ 
\hline 
\textit{Planck} high-$\ell$ TT, TE, EE & 588.29 & 619.76 \\ 
\textit{Planck} low-$\ell$ TT & 22.50 & 23.08 \\ 
\textit{Planck} low-$\ell$ EE & 396.99 & 408.21 \\ 
\textit{Planck} lensing & 9.19 & 21.66 \\ 
BAO low-$z$ & 1.75 & 0.67 \\ 
BAO high-$z$ & 3.47 & 12.55 \\ 
SH0ES & 16.96 & 23.86 \\ 
\hline
Total $\chi^2_\text{min}$ & 1039.15 & 1109.79\\ 
\hline \hline 
\end{tabular}
\caption{\label{tab: chi^2} The best-fit $\chi^2$ per experiment for the standard $\Lambda$CDM model and the AQ model with $\beta$=12 and $\mu=3.5$ Mpc$^{-1/2}$. The BAO low-$z$ and high-$z$ datasets correspond to $0.1<z<0.15$ and $0.38<z<0.61$ respectively. Constraints are based on the CMB, BAO, and R19 datasets. }
\end{table}

\begin{figure*}[t]
\includegraphics[width=\textwidth]{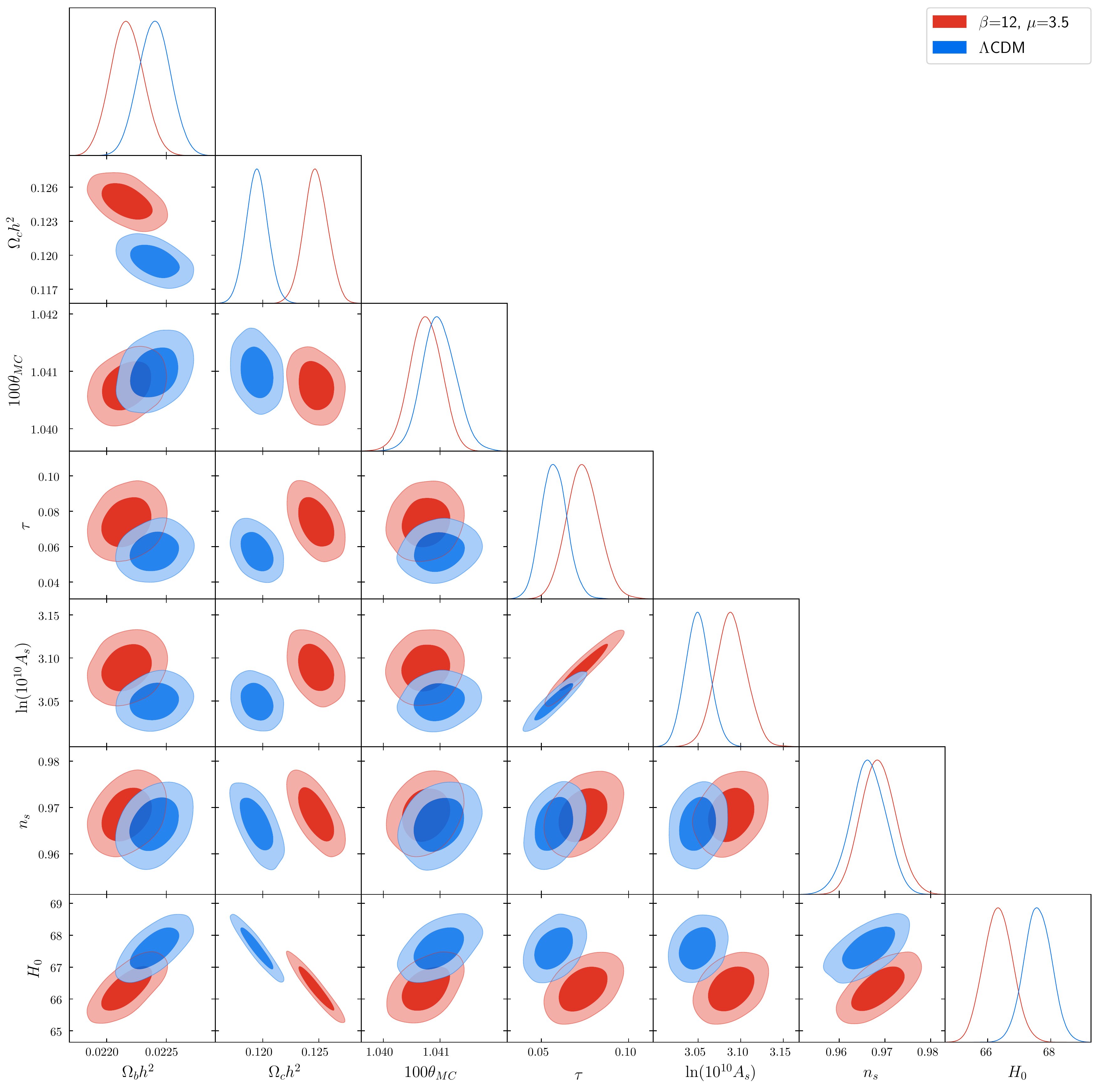}
\caption{\label{fig: P18+BAO} Posterior distributions of the AQ model with $\beta=12$, and $\mu=3.5$ Mpc$^{-1/2}$ (red) and the $\Lambda$CDM model (blue) based on CMB and BAO datasets. The darker inner (lighter outer) regions correspond to 1$\sigma$ (2$\sigma$) confidence intervals.}
\end{figure*}

\begin{figure*}[t]
\includegraphics[width=\textwidth]{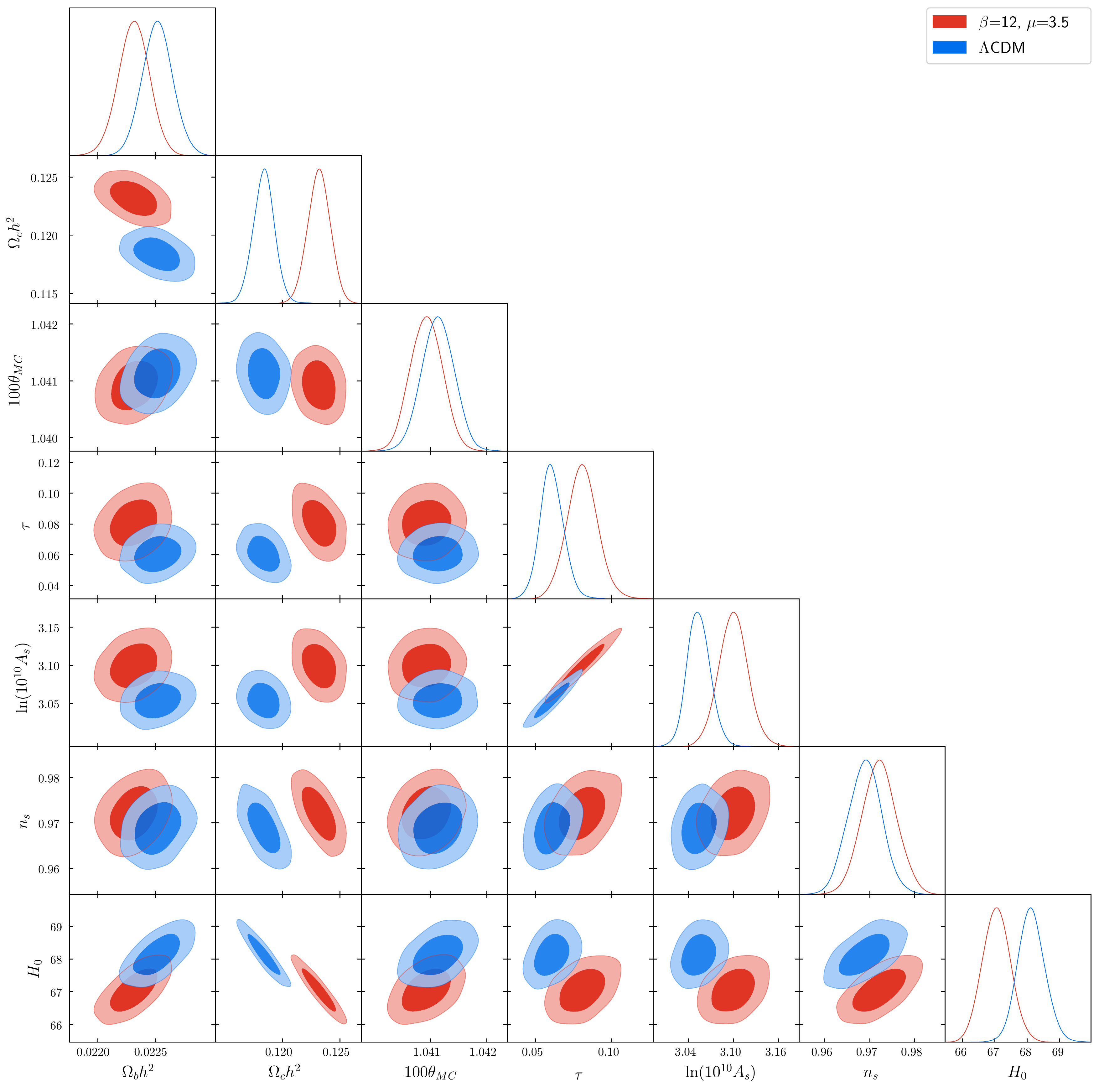}
\caption{\label{fig: P18+BAO+R19} Posterior distributions of the AQ model with $\beta=12$, and $\mu=3.5$ Mpc$^{-1/2}$ (red) and the $\Lambda$CDM model (blue) based on CMB, BAO, and R19 datasets. The darker inner (lighter outer) regions correspond to 1$\sigma$ (2$\sigma$) confidence intervals.}
\end{figure*}

In a series of tables we show the constraints on cosmological parameters in $\Lambda$CDM and the AQ model with fixed model parameters utilizing the CMB and BAO datasets (Table \ref{tab: CMB+BAO}), and the CMB, BAO, and R19 datasets (Table \ref{tab: CMB+BAO+R19}). We show the best-fit $\chi^2$ for individual experiments in these models in Table \ref{tab: chi^2}. The posterior distributions for the relevant parameters are shown in Fig.~\ref{fig: P18+BAO} for the CMB and BAO datasets, and in Fig.~\ref{fig: P18+BAO+R19} for the CMB, BAO and R19 datasets. 

The inclusion of more cosmological data does not change the conclusions made in Sec. \ref{sec: fixed beta and mu}. The parameter changes we saw for \textit{Planck} 2018 data alone shown in Table \ref{tab: P18 constraints} are still present with the inclusion of more datasets. The difference between the best-fit value of $H_0$ in $\Lambda$CDM and the AQ model is narrowed when the BAO and R19 datasets are considered. However, the worse overall fits to the data in the AQ model with fixed model parameters, tell us that this comes at a price. In particular, the AQ model provides a worse fit to the SH0ES likelihood than $\Lambda$CDM, as shown in Table \ref{tab: chi^2}, providing further proof that the effect of the scaling behavior on perturbations in this model are too great to overcome to resolve the $H_0$ tension.

\section{Extended Results for Free Model Parameters}
\label{sec: extended results free beta mu}

\begin{table*}[t]
\begin{tabular}{| c | c | c |}
\hline \hline
 Parameter              & $\Lambda$CDM & $\beta$, $\mu$ free \\ \hline \hline
 $100 \omega_b$         & 2.140 (2.142) $\pm$ 0.015 & 2.097 (2.079) $\pm$ 0.020 \\ 
 $\omega_c$             & 0.1235 (0.1236) $\pm$ 0.0015 & 0.1252 (0.1256)$^{+0.0018}_{-0.0021}$ \\ 
 $100 \theta_s$         & 1.04031 (1.04031) $\pm$ 0.00030 & 1.04024 (1.04032) $\pm$ 0.00032 \\
 $\tau$                 & 0.0459 (0.0473)$^{+0.0083}_{-0.0065}$ & 0.0448 (0.0444)$^{+0.0086}_{-0.0073}$ \\ 
 $\ln (10^{10} A_s)$    & 3.026 (3.030)$^{+0.017}_{-0.014}$ & 3.018 (3.015)$^{+0.019}_{-0.016}$ \\ 
 $n_s$                  & 0.9510 (0.9514)$\pm$ 0.0045 & 0.9486 (0.9491) $\pm$ 0.0052 \\
 $\beta$                & - & 13.3 (10.0)$^{+1.6}_{-3.5}$  \\
 $\mu$ [Mpc$^{-1/2}$]   & - & 1.52 (1.35)$^{0.23}_{0.51}$ \\
 \hline
 $H_0$ [km/s/Mpc]       & 65.17 (65.12) $\pm$ 0.064 & 62.5 (60.57)$^{+1.5}_{-1.1}$ \\ 
 $S_8$                  & 0.870 (0.873) $\pm$ 0.019 & 0.836 (0.8216) $\pm$ 0.026 \\
 \hline
 Total $\chi^2_\text{min}$ & 1575.67 &  1560.80 \\
 \hline \hline 
\end{tabular}
\caption{\label{tab: no lensing constraints} The mean (best-fit) $\pm 1\sigma$ error of the cosmological parameters in the AQ model with free $\beta$ and $\mu$ from our run using only \textit{Planck} high-$\ell$ TT,TT,EE, and low-$\ell$ TT and EE data (i.e. no lensing).}
\end{table*}

\begin{figure*}[t]
\includegraphics[width=\textwidth]{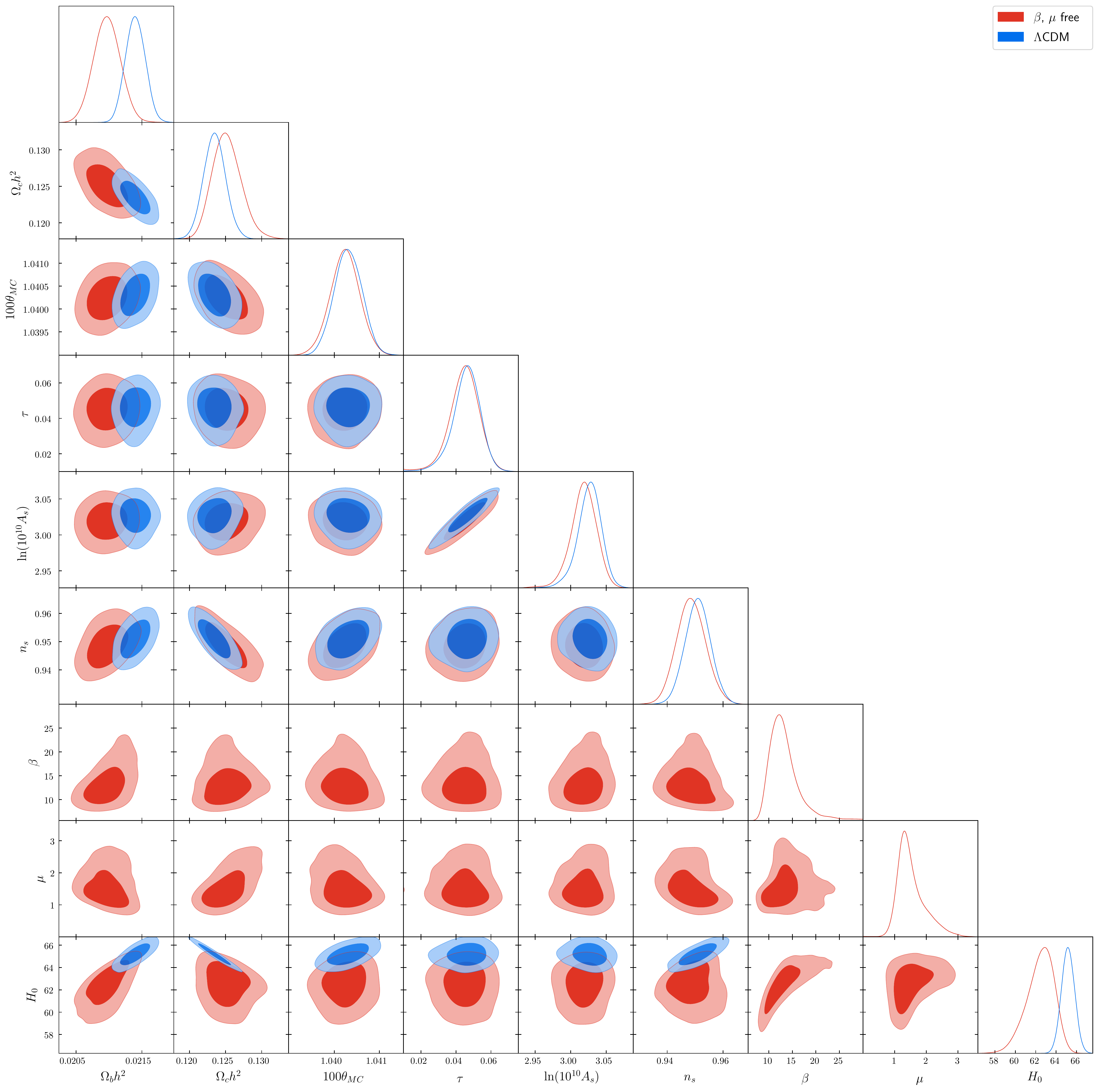}
\caption{\label{fig: no lensing tri-plot} Posterior distributions for the AQ model with free $\beta$ and $\mu$ (red) and the $\Lambda$CDM model (blue) for the \textit{Planck} high-$\ell$ TT,TT,EE, and low-$\ell$ TT and EE data. The darker inner (lighter outer) regions correspond to 1$\sigma$ (2$\sigma$) confidence intervals.}
\end{figure*}

In this Appendix we present the results of our MCMC analysis on $\Lambda$CDM and the AQ model with free model parameters utilizing the \texttt{TTTEEE Plik lite} high-$\ell$, and \texttt{TT} and \texttt{EE} low-$\ell$ likelihoods. We give the constraints on cosmological parameters in Table \ref{tab: no lensing constraints} and the posterior distributions for all parameters in Fig.~\ref{fig: no lensing tri-plot}. 

These constraints show that with the effect of CMB lensing turned off, an AQ scaling field which becomes dynamical after recombination provides a statistically better fit to the \textit{Planck} temperature and polarization data than $\Lambda$CDM with $\Delta\chi^2_\text{min}= -14.87$ as seen in Table \ref{tab: no lensing constraints}. This is likely because the AQ field lowers the depth of gravitational potentials, smoothing the CMB spectrum which mimics the effect of gravitational lensing. Since the introduction of lensing results in a much better fit to \textit{Planck} measurements in the $\Lambda$CDM model, the AQ field brings the AQ model into better agreement with \textit{Planck} measurements by mimicking this effect. However the resulting best-fit value of the Hubble constant is $H_0=60.57$ km/s/Mpc. This result, combined with the ``$S8$-tension'' and the $A_L$ anomaly present in \textit{Planck} 2018 data, suggest the need for a more general analysis of cosmological data, with relaxed assumptions of dark energy, lensing, and expansion history.

\bibliography{main}

\end{document}